\documentclass[11pt,twoside]{article}
\usepackage{asp2010}

\resetcounters

\begin{document}

\title{SPICA as a probe of cosmic reionization}
\author{Erik~Zackrisson$^1$ and Akio~K.~Inoue$^2$
\affil{$^1$The Oskar Klein Centre, Department of Astronomy, Stockholm University, Sweden}
\affil{$^2$College of General Education, Osaka Sangyo University, Japan}}

\begin{abstract}
Current data indicate that the reionization of the Universe was complete by redshift $z\approx 6$--7, and while the sources responsible for this process have yet to be identified, star-forming galaxies are often considered the most likely candidates. However, the contribution from $z\gtrsim 6$ galaxies to cosmic reionization critically depends on the fraction of ionizing (Lyman continuum, LyC) photons escaping from these objects and into the intergalactic medium. At $z \lesssim 4$, the escaping LyC flux can be measured directly, but the opacity of the neutral intergalactic medium precludes such measurements at higher redshifts. In a recent paper, we argue that since the LyC escape fraction regulates the contribution of nebular emission to the rest-frame optical/UV spectra of galaxies, the James Webb Space Telescope should be able to indirectly assess the LyC escape fraction for galaxies at $z\approx 6$--9. JWST can, on the other hand, not constrain the fraction of LyC photons directly absorbed by dust, and this is where SPICA comes in. The dust continuum emission from gravitationally lensed LyC-leakers at $z\approx 6$ may in principle be detectable with SPICA, thereby constraining the level of LyC extinction in these objects.
\end{abstract}

\section{Introduction}
Several independent lines of evidence indicate that cosmic reionization was complete by $z\approx 6$--7 \citep[e.g.][]{Fan06,Zahn12}, and while star-forming galaxies at $z\gtrsim 6$ may well be numerous enough to produce the required LyC photons \citep[e.g.][]{Finkelstein12,Alvarez12,Mitra13,Robertson13}, this scenario requires that LyC photons can escape from these galaxies and into the intergalactic medium (IGM). LyC leakage from galaxies can be measured directly at $z\lesssim 4$, but the opacity of the increasingly neutral IGM prevents such measurements at higher redshifts \citep{Inoue&Iwata}. In a recent paper \citep{Zackrisson13}, we argue that since the LyC escape fraction $f_\mathrm{esc}$ regulates the impact of nebular emission on the spectra of galaxies, it will nonetheless be possible to indirectly probe LyC leakage well into the reionization epoch. Spectroscopy with the James Webb Space Telescope  (JWST, scheduled for launch in 2018) should be able to provide a rough handle on $f_\mathrm{esc}$ for galaxies at $z\approx 6$--9, but since JWST cannot probe dust emission from such objects, there will be lingering uncertainties regarding the fraction of LyC flux lost due to dust absorption. Here, we explain how SPICA can contribute to this field by measuring the dust continuum emission from gravitationally galaxies at $z\approx 6$.

\section{Dust emission and the role of SPICA}
Comparisons between ultraviolet, optical and far-infrared (FIR) star-formation indicators suggest that a substantial fraction of the LyC photons produced in local star-forming galaxies are directly absorbed by dust \citep{Inoue01,Hirashita03}. If this effect could be quantified also for reionization-epoch galaxies, this would greatly simplify the task of pinning down the LyC escape fractions of such objects \citep{Zackrisson13}. Since the dust heated by LyC photons re-emits the absorbed energy at rest-frame FIR wavelengths, this requires some handle on the FIR luminosities of galaxies at $z\gtrsim 6$. 

The dust emission of $z\approx 6$--9 galaxies is predicted to peak at a few hundred $\mu$m \citep{Takeuchi05}, and can in principle be probed with the Atacama Large Millimeter/sub-millimeter Array (ALMA), or upcoming telescopes like the Space Infrared Telescope for Cosmology and Astrophysics (SPICA) and the Cerro Chajnantor Atacama Telescope (CCAT).  However, the low intrinsic luminosities of these galaxies makes this an extremely challenging task. Gravitational lensing by foreground galaxy clusters can boost the fluxes of galaxies in the reionization epoch by a factor of up to $\mu\sim 100$ \citep[e.g.][]{Zackrisson12,Bradley13}, and the best targets for this endeavour are therefore likely to lie in strongly lensed fields.

\begin{figure}[!ht]
\plotone{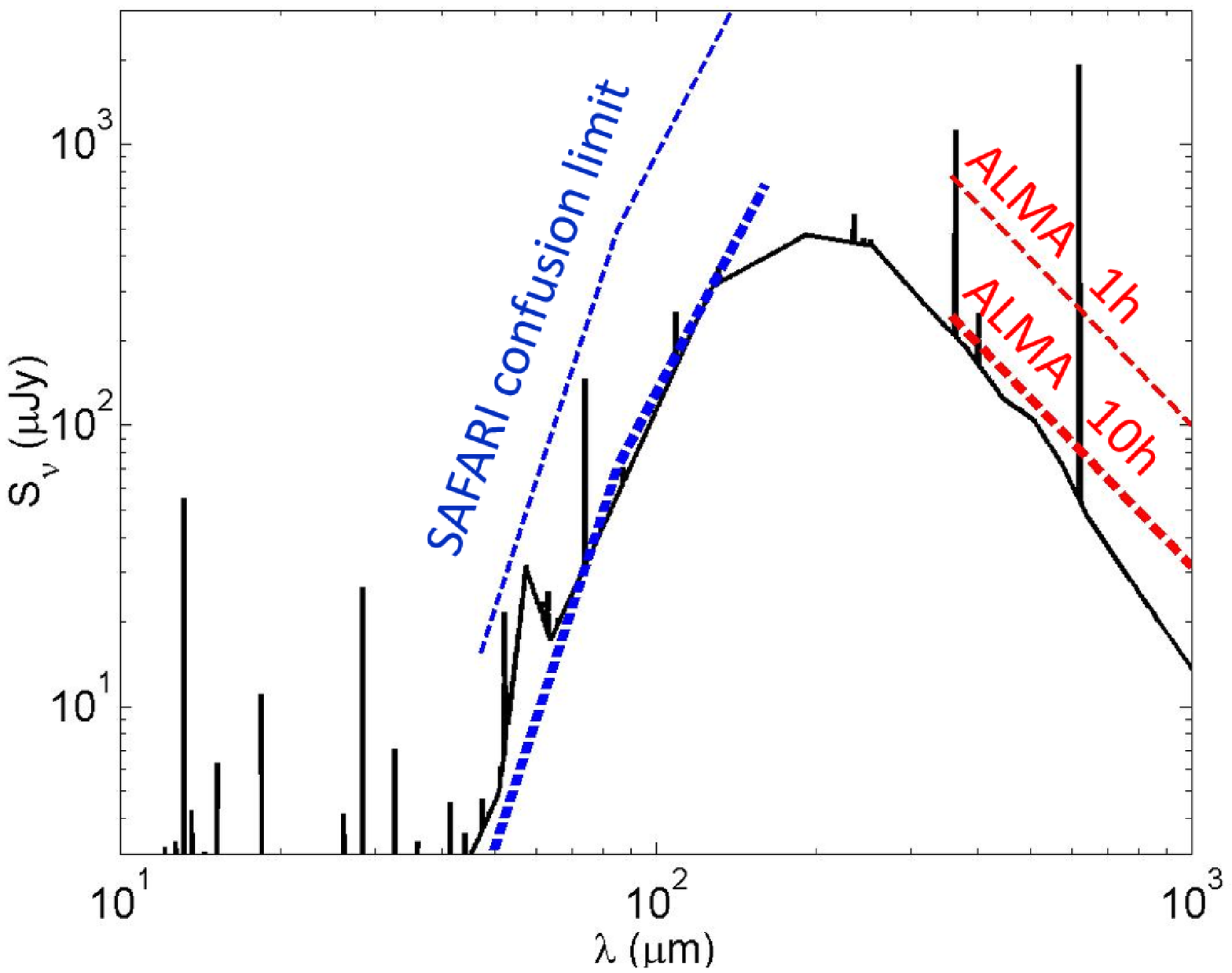}
\caption{The dust emission peak of a gravitationally lensed galaxy at $z=6$ (magnification $\mu=100$). The dust bump predicted by \citet{Takeuchi05} for a $z=6$ dwarf galaxy with star formation rate SFR = 10 $M_\odot\ \mathrm{yr}^{-1}$ has here been superposed on a matching {\it Yggdrasil} stellar + nebular galaxy spectrum \citep{Zackrisson11}. The SPICA confusion limit and the ALMA 1h and 10h detection limits are marked by dashed lines. In this case, detection of the high-energy part of dust emission bump requires photometry with SPICA at flux levels a factor of several below the confusion limit (the lower, thick dashed SPICA line marks the threshold reached a factor of 7 below the formal confusion limit). This may be possible by taking advantage of the deep, high-resolution observations at both shorter (JWST) and longer (ALMA) wavelengths. ALMA may in principle detect the low-energy part of the dust bump, but this would requires integration times of more than 10 h.}
\label{fig1}
\end{figure}

In Figure~\ref{fig1}, we plot the predicted dust continuum bump for a SFR = 10 $M_\odot\ \mathrm{yr}^{-1}$ dwarf galaxy \citep{Takeuchi05}, superposed on an {\it Yggdrasil} \citep{Zackrisson11} model for the stellar and nebular (line and continuum) components of the spectrum. The object is assumed to be at $z=6$ and to be gravitationally magnified by a factor of $\mu = 100$.  At this redshift, i.e.~ at the end of the reionization epoch, the peak of the dust continuum bump falls right between the SPICA/SAFARI and ALMA detection windows, which makes estimates of the total luminosity in the bump very tricky. ALMA can in principle detect the low-energy (high-wavelength) part of the bump, but this requires integration times of more than 10 hours. SPICA/SAFARI is better suited to measure the high-energy (low-wavelength) part of the bump, but is confusion limited at these wavelengths. SAFARI is able to reach its $\approx 50$--160 $\mu$m confusion limits in two minutes or less, but to reach the dust continuum, one needs to push the detection limit a factor of several below this limit. We suggest that this may be accomplished through the use of auxiliary, high-resolution data from other wavelength bands (e.g. obtained with JWST or ALMA) to subtract off the estimated flux contribution from nearby objects in the SPICA bands. However, detailed simulations are required to determine the exact limits of this technique. While CCAT may be able to observe at wavelengths down to 200 $\mu$m, its sensitivity is insufficient to reach the peak of the dust continuum bump for $z=6$ galaxies of this type, as this would require integration times of much more than 100 hours.

\section{Summary}
An important step in the quest to identify the sources responsible for the reionization of the Universe is to determine the typical LyC escape fractions of galaxies at $z\gtrsim 6$. Since the LyC escape fraction regulates the ratio of nebular emission to direct star light in the rest-frame ultraviolet/optical part of the spectrum, spectroscopy with JWST should allow LyC escape fractions of individual galaxies at $z\approx 6$--9 to be constrained. However, JWST by itself cannot assess the fraction of LyC photons directly absorbed by dust. At $z\approx 6$, this effect may instead be investigated through SPICA observations of the brightest, gravitationally lensed galaxies. Observations of this type are, however, extremely challenging, and require photometry at fluxes a factor of several below the formal SPICA confusion limits at 50--160 $\mu$m. However, this can possibly be accomplished through the use of auxiliary, high-resolution observations from other wavelength bands.

\acknowledgements 
E.Z acknowledges funding from the Swedish National Space Board, the Swedish Research Council and the Wenner-Gren Foundations. A.K.I. acknowledges funding from the Ministry of Education, Culture, Sports, Science, and Technology (MEXT) of Japan (KAKENHI: 23684010).

\bibliography{zackrisson}

\end{document}